\documentclass[preprint]{aastex62}
\usepackage{mathrsfs}
\usepackage{graphicx}
\usepackage{multirow}
\usepackage{amsmath}
\usepackage{threeparttable}
\usepackage{float} 
\usepackage{epstopdf}
\received{}
\revised{}
\accepted{}
\submitjournal{ApJ}

\shorttitle{Constraining the Diffusion Coefficient within the TeV Nebula}
\shortauthors{Bao et al.}

\begin{document}
\title{\Large{\textbf{On the Gamma-ray Nebula of Vela Pulsar. I. Very Slow Diffusion of Energetic Electrons within the TeV Nebula}}}

\correspondingauthor{Siming Liu}
\email{liusm@pmo.ac.cn}

\correspondingauthor{Yang Chen}
\email{ygchen@nju.edu.cn}

\author{Yiwei Bao}
\affil{Department of Astronomy, Nanjing University, 163 Xianlin Avenue, Nanjing 210023, China}

\author{Siming Liu}
\affil{Key Laboratory of Dark Matter and Space Astronomy, Purple Mountain Observatory, Chinese Academy of Sciences, Nanjing 210034, China}
\affil{School of Astronomy and Space Science, University of Science and Technology of China, Hefei 230026, China}

\author{Yang Chen}
\affil{Department of Astronomy, Nanjing University, 163 Xianlin Avenue, Nanjing 210023, China}
\affiliation{Key Laboratory of Modern Astronomy and Astrophysics, Nanjing University, Ministry of Education, Nanjing, China}




\begin{abstract}
High-energy particle transport in pulsar wind nebulae (PWNe) plays an essential role in explaining the characteristics revealed in multiwavelength observations. In this paper, {the TeV-gamma-ray-emitting electrons in the Vela X PWN are approximated to be injected impulsively when the cocoon is formed due to the interaction between the SNR reverse shock and the PWN. By solving the diffusion-loss equation analytically, we reproduce the broadband spectral energy distribution and surface brightness profile simultaneously. The diffusion coefficient of TeV electrons and positrons, which is well constrained by the spectral and spatial properties of the TeV nebula, is thus determined to be $1 \times 10^{26}$\,cm$^{2}$\,s$^{-1}$ for 10\,TeV electrons and positrons. This coefficient is} more than three orders of magnitude lower than that in the interstellar medium, in agreement with a constraint recently obtained from HAWC observations of a TeV nebula associated with the Geminga pulsar. These results suggest that slow diffusion of high-energy particles might be common in PWNe. 
\end{abstract}


\section{INTRODUCTION}\label{sec:intr}
The unexpected excess of high-energy positrons ($\geq 10$ GeV) observed by PAMELA \citep{2013PhRvL.111h1102A}, AMS-02 \citep{2014PhRvL.113l1101A}, VERITAS \citep{2015ICRC...34..411S}, \textit{Fermi}-LAT \citep{2012PhRvL.108a1103A,2017PhRvD..95h2007A}, and DAMPE \citep{2017Natur.552...63D} has been explored extensively by many researchers, yet its origin is still a matter of debate. While dark matter particles are believed to be a possible positron source, pulsars and pulsar wind nebulae (PWNe) are more natural astrophysical contributors owing to their high pair production capability \citep{2009JCAP...01..025H}. Pulsars inside PWNe produce ultra-relativistic wind of electron-positron pairs (hereafter, ``electrons'' refers to both electrons and positrons). A shock forms as the wind interacts with slow-moving supernova ejecta, and the electrons are thermalized and/or reaccelerated by the shock \citep{2006ARA&A..44...17G}. The resulting high-energy electrons may escape from PWNe, and propagate in the nearby interstellar medium (ISM). Considering radiative energy loss experienced by high-energy electrons while propagating, TeV positron sources must be located within hundreds of parsecs from Earth {if they are to contribute to the positron excess.}

However, this scenario is questioned by the recent HAWC's detection of extended TeV $\gamma$-rays from the Geminga PWN and PSR B0656+14. Although this observation confirms the existence of TeV electrons in PWNe, the diffusion coefficients, constrained by the $\gamma$-ray surface brightness profile, are too small to account for the positron flux observed at Earth \citep{2017Sci...358..911A}. {Two-zone diffusion models seem to partially remedy this issue \citep{2017PhRvD..96j3013H,2018ApJ...863...30F}.} Nevertheless, the local PWNe origin of the positron excess is oppugned by non-detection of the Geminga PWN in \textit{Fermi} observations \citep{2018arXiv181010928X}, {and the nature of positron excess is still under debate \citep{2019MNRAS.tmp..670S}.}

The Vela pulsar, lying at a distance of 287\,pc to Earth \citep{2003ApJ...596.1137D}, is one of the positron contributor candidates  in the TeV energy range. Thanks to its proximity, the Vela X PWN is spatially resolved in multiwavelength observations from the radio to $\gamma$-rays. The extended radio nebula (ERN) of Vela X, which is spatially correlated with the GeV nebula seen by \textit{Fermi}, is much larger than the {X-ray ``cocoon" \citep{1995Natur.375...40M,1997ApJ...475..224F}. A counterpart of the cocoon was later revealed in TeV $\gamma$-rays \citep{2006A&A...448L..43A}.} Although there are extensive observational and theoretical studies, the formation of the complex frequency-dependent morphology {of the PWN} remains to be addressed \citep{2018A&A...617A..78T}. Two electron components have been assumed to explain the frequency dependent morphology \citep{2008ApJ...689L.125D}. However, challenges arise when TeV emission beyond the cocoon {(extended to $\sim 1.2^\circ$) is detected \citep{2012A&A...548A..38A,2017SSRv..207..175R}. We hereafter refer to the TeV-$\gamma$-ray-emitting plasma as the TeV nebula.} \citet{2011ApJ...743L...7H} argued that diffusive escape of high-energy electrons must be taken into consideration to explain the soft GeV spectrum of the ERN while the TeV {nebula} formed over the past few hundreds of years, so that high-energy electrons are still trapped in a cocoon structure. The escaped electrons, if diffused to Earth, should contribute to the local cosmic ray positron spectrum, suggesting that the Vela X PWN may be one of the positron contributors \citep{2011ApJ...743L...7H}. {However, AMS-02 measurement shows that the positron fraction saturates at $\sim 300$ GeV and likely drops above $\sim 500$ GeV \citep{2018arXiv181107551R,2019PhRvL.122d1102A}}, disfavoring the pulsars and PWNe as the origin of the positron excess.

\citet{2018A&A...617A..78T} found that the TeV {nebula of the Vela X PWN} has a very hard spectrum above 100 GeV. This spectrum {suggests that the emission is dominated by electrons at tens of TeV}. Such a distribution is a natural consequence of a hard electron spectrum with a power-law index of less than 2 being subjected to synchrotron and IC radiative energy losses. The {spatial offset} of the cocoon from pulsar suggests that it was likely caused by interaction with the reverse shock of the supernova remnant (SNR) thousands of years ago \citep{2018ApJ...865...86S}. We therefore propose that the {TeV nebula} formed via {impulsive} injection of a hard power-law distribution of energetic electrons before the PWN was displaced by interaction with the reverse shock several thousands year ago. The consequent parallel diffusion and radiative energy loss of these high-energy particles in a large-scale magnetic field can naturally account for the spatial and spectral properties of the TeV {nebula} \citep{2003MNRAS.343..116D}. 

Based on the escape model proposed by \citet{2011ApJ...743L...7H}, \citet{2018ApJ...866..143H} argued that diffusion coefficient in the Vela X PWN should be less than $10^{28}$ cm$^2$ s$^{-1}$ for 10\,TeV electrons to avoid excessive cosmic ray electron flux above 10\,TeV. Here we suggest that the extended TeV emission beyond the X-ray cocoon arises from diffusion of high-energy electrons and use the spatial and spectral properties of the TeV {nebula} to constrain the diffusion coefficient in this nebula directly. The electrons are injected as a power-law function of energy, {and the diffusion-loss equation can be solved analytically}. Since the effective magnetic field strength for radiative energy loss can be constrained by the high-energy cutoff of the TeV spectrum and the $\gamma$-ray emission is dominated by {the electrons at tens of TeV}, the diffusion coefficient is essentially the only key parameter one can adjust to fit the TeV $\gamma$-ray radial brightness profile and the spectra of the inner region ($0^\circ$--$0.8^\circ$, {slightly larger than the X-ray cocoon}) and an outer ring ($0.8^\circ$--$1.2^\circ$, {diffused emission beyond the cocoon; \citeauthor{2012A&A...548A..38A}\,2012}). Our model is described in \S 2. The application of this model to the TeV nebula of the Vela pulsar is presented in \S 3. The discussion and conclusions are provided in \S 4. {In a follow-up paper (\citeauthor{paperII}\ 2019, hereafter Paper II), we explain the spectrum of the ERN and the associated GeV $\gamma$-ray emission.}

\section{MODEL DESCRIPTION}
We assume the TeV surface brightness profile stems from the diffusion {and radiation losses} of high-energy electrons. For simplicity, a simple diffusion-loss model is adopted to constrain the diffusion coefficient within the TeV nebula of Vela pulsar. The transport of electrons can be described by
\begin{eqnarray}
\label{eq:dif}
\frac{\partial}{\partial t}f(\gamma,r,t)=\frac{D(\gamma)}{r^{2}}\frac{\partial}{\partial r}r^{2}\frac{\partial}{\partial r}f(\gamma,r,t)+\frac{\partial}{\partial \gamma}\left(Pf\right)+Q_\textup{inj}(\gamma,t),
\end{eqnarray}
{where $f(\gamma,r,t)$ is the electron distribution function}, $r$ is the radial distance to the center of the TeV nebula, $D(\gamma)=D_0\left(\gamma/\gamma_{_\textup{10\,TeV}}\right)^\delta$ is an energy-dependent diffusion coefficient (with $\gamma_{_\textup{10\,TeV}}$ the Lorentz factor of 10\,TeV electrons), $P$ is the radiation energy loss rate, and $Q_\textup{inj}$ is the electrons' injection rate associated with the pulsar's spin-down. {Since the slow-diffusion regions ($\sim$ 30--50\,pc in radii) in the two-zone diffusion models \citep{2018ApJ...863...30F,2018ApJ...866..143H} are apparently larger than the Vela SNR ($\sim 20$\,pc in radius), we assume a spatially independent diffusion coefficient. The $\gamma$-ray spectral indices are almost a constant throughout the cocoon \citep{2012A&A...548A..38A}, therefore the diffusion coefficient is expected to have a weak energy dependence; we thus fix $\delta=1/3$ (Kolmogorov diffusion, as is assumed in \citeauthor{2017Sci...358..911A}\ 2017)}. {Although a broken power law is usually used to explain the broadband spectral energy distribution (SED) of PWNe, a single power law is sufficient to explain the broadband spectrum of the Vela X PWN (see Paper II), therefore the injection rate is assumed to be}  
\begin{equation}
Q_\textup{inj}(\gamma,t)=Q_0\delta(t-\tau_\textup{s})\left(\frac{\gamma}{10^7}\right)^{-\alpha},
\label{eq:Q}
\end{equation}
where $Q_0$ is the injection {constant} for the electrons with $\gamma=10^7$. {Here we assume that most of plasma injected before the reverse shock-PWN interaction is compressed into the ERN, and approximate the injection to be impulsive for the following reasons. The total energy in the cocoon is estimated to be $E_\textup{cocoon}\sim 1.5 \times 10^{46}$\,erg \citep{2010ApJ...713..146A}, much lower than the energy in the ERN ($5 \times 10^{48}$\,erg; \citeauthor{2010ApJ...713..146A}\,2010). Hydrodynamic simulations have revealed that the cocoon was formed $\sim 4$\,kyr ago \citep{2018ApJ...865...86S}, when the SNR reverse shock disrupted the PWN, creating a tail (the cocoon) to the south, which contains a small fraction of fresh plasma released by the pulsar. Meanwhile, the relatively faint TeV emission and bright X-ray emission in the very vicinity of the Vela pulsar imply a strong magnetic field formed after the reverse shock interacted with the PWN \citep[see, e.g.,][]{2011ApJ...743L...7H} that can cool the electrons (injected after the X-ray cocoon was created) to low energy in a short time ($\sim 10^2(B/400\ \mu\textup{G})^{-2}(E/700\ \rm GeV)^{-1} $\,yr). Therefore, injection after the cocoon was formed does not contribute to TeV electrons. The cooling timescale of 70\,TeV electrons is $\sim 4.4 \times 10^3(B/6\mu\textup{G})^{-2}$\,yr \citep{2011ApJ...743L...7H}, hence the very high-energy electrons injected $\sim 4$\,kyr ago have been cooled down, giving rise to the observed TeV spectral cutoff.}

Given the very hard GeV spectrum of the TeV nebula \citep{2012A&A...548A..38A,2018A&A...617A..78T}, $\alpha$ needs to be less than 2, giving rise to an {energy} distribution piling up where the radiative cooling timescale is equal to the age of the injected electrons (see \autoref{fig:elec}). To get the structure of the nebula, we assume that the nebula is spherical symmetric\footnote{ 
	Although the X-ray cocoon and its TeV counterpart is hardly spherically symmetric in the central region ($\sim 0.8^\circ$ in length), spherical symmetry could be a good approximation for the spatially integrated spectra and the larger extension of TeV emission out to $2^\circ$ in radius. On the other hand, the asymmetry in the central region would only affect the first two data points in the surface brightness profile. Also, the radial TeV brightness profile is given from circular annuli \citep{2012A&A...548A..38A}, so spherical symmetry could be a convenient technical treatment to fit the data. Moreover, the diffusion coefficient is derived from fitting the relative surface brightness, therefore the absolute magnitude of the injected energy is of less importance. The asymmetry does not affect our constraint on the diffusion coefficient obtained. The SED, however, depends on the total number of the $\gamma$-ray-emitting electrons injected in the TeV nebula after interaction with the reverse shock, which can be adjusted by changing the injection constant $Q_0$}. Here we only consider the synchrotron and inverse Compton (IC) losses, {and the loss rate can be} described by $d\gamma/dt=-p_2\gamma^2${, where $p_2=3.23 \times 10^{-20} (B/5\,\mu G)^{2}\ \textup{s}^{-1}$. Following the procedure described in \citet{1995PhRvD..52.3265A}, we define $\mathscr{F}=(d\gamma/dt)rf(\gamma, r, t)$, hence \autoref{eq:dif} can be written as}
\begin{equation}
\frac{\partial\mathscr{F}}{\partial z}=D_1(z)\frac{\partial^2\mathscr{F}}{\partial r^2}
\end{equation}
(see \citeauthor{1995PhRvD..52.3265A}\,1995 for the definition of $z$ and $D_1(z)$); consequently, the electron distribution function {$f(\gamma, r, t)$ can be obtained:} 
\begin{eqnarray}
\label{eq:solution}
f(\gamma,r,t)=\left \{
\begin{array}{ll}
\frac{\gamma^2_tQ_\textup{inj}(\gamma_t,t)}{\gamma^2\pi^{3/2}r^3_\textup{dif}}\exp\left({-\frac{r^2}{r^2_\textup{dif}}}\right) &\ \ \gamma \le \gamma_{_\textup{max}},\\
0                                                                                  & \ \ \gamma > \gamma_{_{\rm max}}.\\
\end{array}  \right .\\
r_\textup{dif}(\gamma,t)=2\sqrt{D(\gamma)t\frac{1-(1-{\gamma}/{\gamma_{_\textup{max}}})^{1-\delta}}{(1-\delta){\gamma}/{\gamma_{_\textup{max}}}}},
\end{eqnarray}
where $\gamma_\textup{max}=1/(p_2t)$, $\gamma_t=\gamma/(1-p_2t\gamma)$ is the initial energy of the electrons that are cooled down to $\gamma$ after time $t$. {\citet{1995PhRvD..52.3265A} showed, $r_\textup{dif}\approx2\sqrt{D(\gamma)t}$ for $\gamma<0.5\gamma_{_{\rm max}}$}. Then we integrate the electron distribution function along the line of sight $l$ (noting that $l^2+R^2=r^2$, with $R$ being the projection distance): 
\begin{equation}\label{eqn:LO}
F_\textup{LS}=\int^\infty_{-\infty} f dl=\int^\infty_{-\infty} \frac{\gamma^2_tQ_\textup{inj}(\gamma_t)}{\gamma^2\pi^{3/2}r^3_\textup{dif}}\exp\left({-\frac{l^2+R^2}{r^2_\textup{dif}}}\right) dl=\frac{\gamma^2_tQ_\textup{inj}(\gamma_t)}{\gamma^2\pi r^2_\textup{dif}}\exp\left({-\frac{R^2}{r^2_\textup{dif}}}\right).
\end{equation} 
According to \citet{2012A&A...548A..38A}, the radial brightness profile is extracted from {annuli} with {the same width of} 12'. {To obtain the $\gamma$-ray fluxes observed now}, $F_\textup{LS}$ is further integrated over each {annulus and pulsar lifetime:}
\begin{equation}
\int^{T_\textup{age}}_0\int_{R_\textup{in}}^{R_\textup{out}} F_\textup{LS} 2\pi R\,dR\,dt=\frac{\gamma^2_t Q_0 \left(\gamma/10^7\right)^{-\alpha}}{\gamma^2}\left[\exp\left({ -\frac{D^2_\textup{in}}{r^2_\textup{dif}(\gamma,T_\textup{age}-\tau_s) }}\right)-\exp\left({ -\frac{D^2_\textup{out}}{r^2_\textup{dif}(\gamma,T_\textup{age}-\tau_s)} }\right)  \right],
\label{eq:solution}
\end{equation}
where $R_\textup{in}$ and $R_\textup{out}$ represent the inner and the outer radius of an {annulus}, respectively, and $T_\textup{age}$ represents the age of the Vela PWN.

\section{Application to Vela X} 
\begin{center}
\begin{figure}[H]
\includegraphics[scale=0.48]{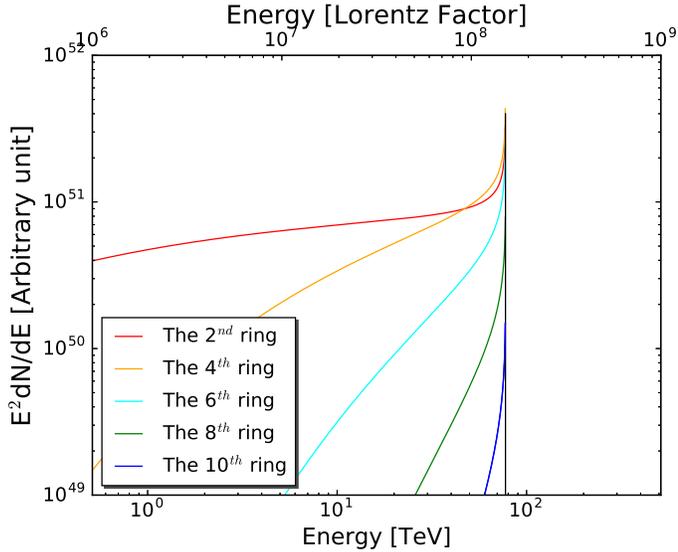}
\caption{Spatially integrated electron distribution function of {annuli} for $\alpha=1.7$. {There are 10 annuli with a width of $12'$.}}
\label{fig:elec}
\end{figure}
\end{center}

The Vela pulsar and its associated SNR have been extensively studied thanks to the pulsar's short distance of 287\,pc obtained using VLBI parallax \citep{2003ApJ...596.1137D}. The age of the pulsar derived from its proper motion is $9000$--$27000$ yr \citep{1995Natur.373..587A}. It is the first H.E.S.S. TeV source showing a prominent high-energy spectral cutoff \citep{2012A&A...548A..38A}. \citet{2006A&A...448L..43A} showed that the magnetic field strength in the X-ray cocoon is $\sim$ 4 $\mu$G based on its X-ray flux. The formation of the X-ray cocoon is attributed to the reverse shock-PWN interaction thousands of years ago \citep{2018ApJ...865...86S}, and the TeV counterpart can be explained by the electron diffusion from the cocoon. We focus on the $\gamma$-ray data and consider the cosmic microwave background radiation (CMB) and the far infrared radiation (FIR) as the background photons for calculation of the $\gamma$-rays via the IC process. We assume blackbody spectrum with temperature of 25\,K and energy density of 0.2 eV cm$^{-3}$ to approximate the FIR radiation field presented in \citet{2008ApJ...682..400P}. We fit the $\gamma$-ray spectra of both the inner and outer regions as well as the surface brightness profile given in \citet{2012A&A...548A..38A}. The model parameters are listed in \autoref{tab:par}, where $d_\textup{Vela}$ is the distance to the Vela pulsar, $B_\textup{eff}$ is the effective magnetic field strength ($B_\textup{eff}^2=B^2+8\pi u_{_{\textup{CMB}}}+ 8\pi u_{_{\textup{FIR}}}$, with $u_{_\textup{CMB}}$ and $u_{_{\textup{FIR}}}$ being the energy density of CMB and FIR photons, respectively). {Because the energy losses are dominated by synchrotron radiation and IC off low-energy photons, the energy losses are assumed to be quadratic in energy. However, the Klein-Nishina effect is incorporated into our calculation of SEDs. Because the electron spectrum is very hard and unbroken below 70\,TeV (see \autoref{fig:elec}), IC off starlight is severely suppressed by the Klein-Nishina effect. Meanwhile, the energy density of starlight is as low as 0.3 eV cm$^{-3}$, with a temperature of $\sim 3000$ K \citep{2008ApJ...682..400P}. The IC off starlight contributes $\sim 1\%$ of the $\gamma$-ray emission and is thus neglected.}

\begin{center}
\begin{figure}[H]
\includegraphics[scale=0.48]{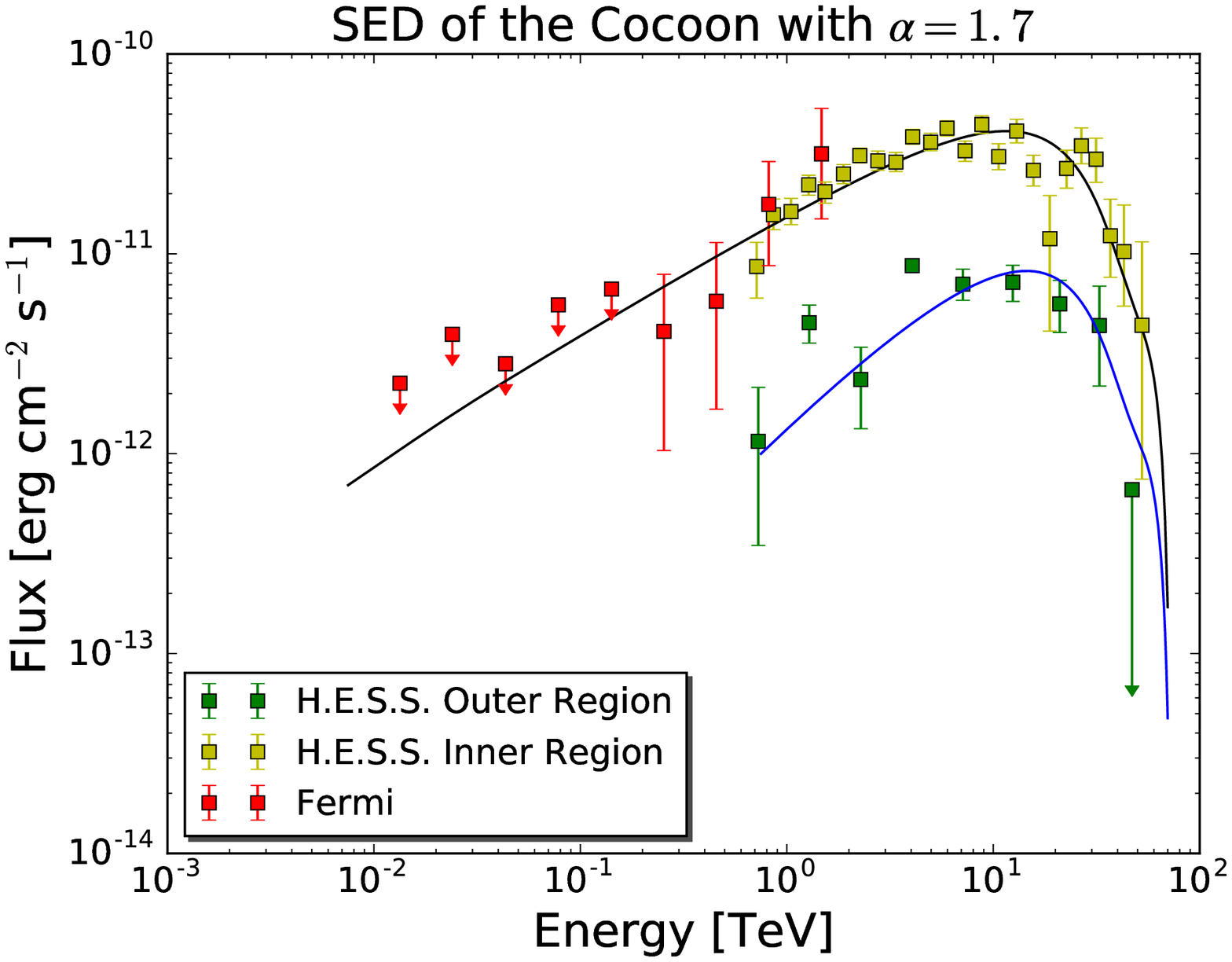}
\includegraphics[scale=0.48]{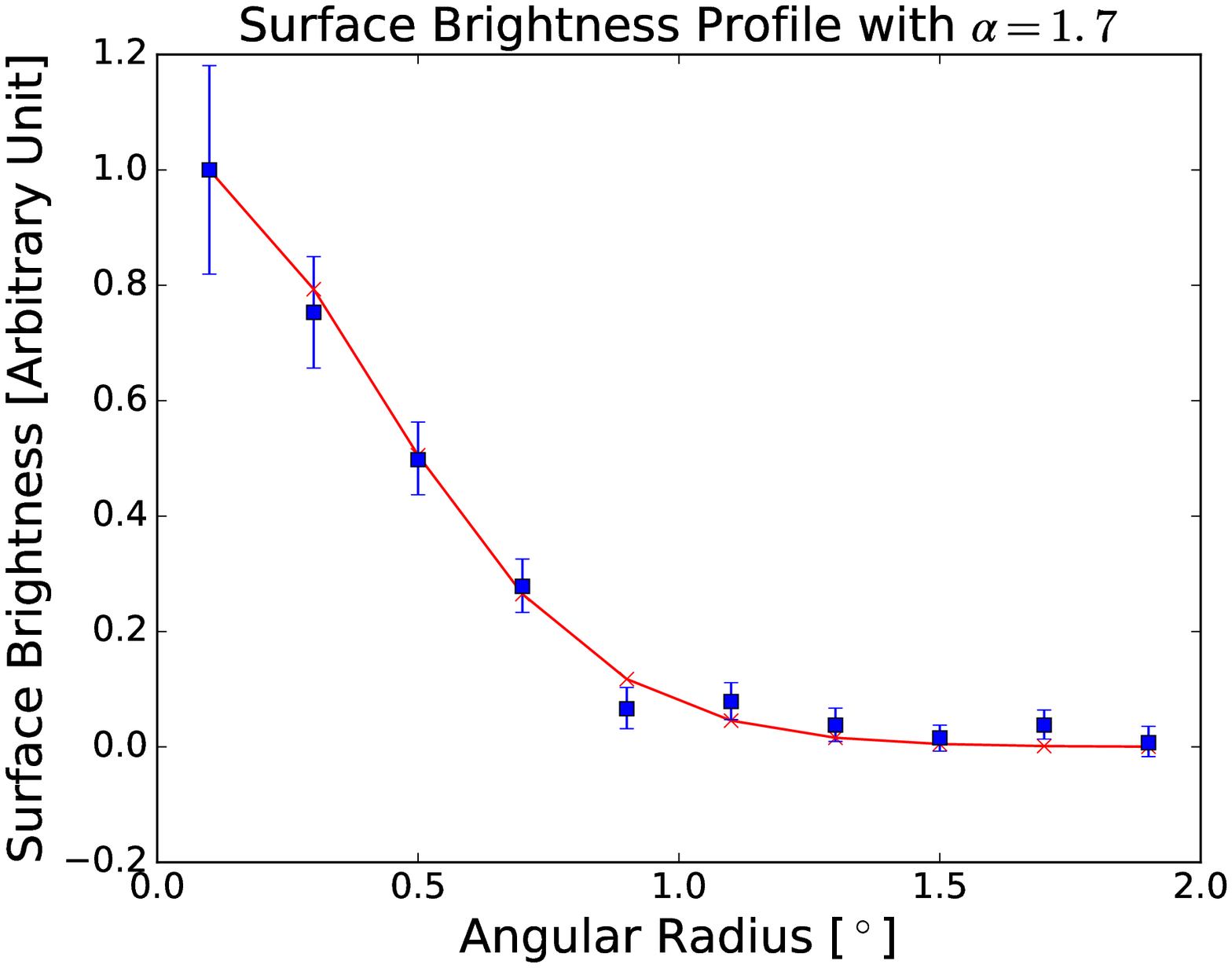}
\caption{Fit to the $\gamma$-ray spectra of the inner and outer regions (left), and the normalized surface brightness profile (right). The H.E.S.S. data are taken from \citet{2012A&A...548A..38A}, {\textit{Fermi} data are from \citet{2018A&A...617A..78T}.}}
\label{fig:Rings}
\end{figure}
\end{center}

\begin{center}
	\begin{figure}[H]
		\includegraphics[scale=0.48]{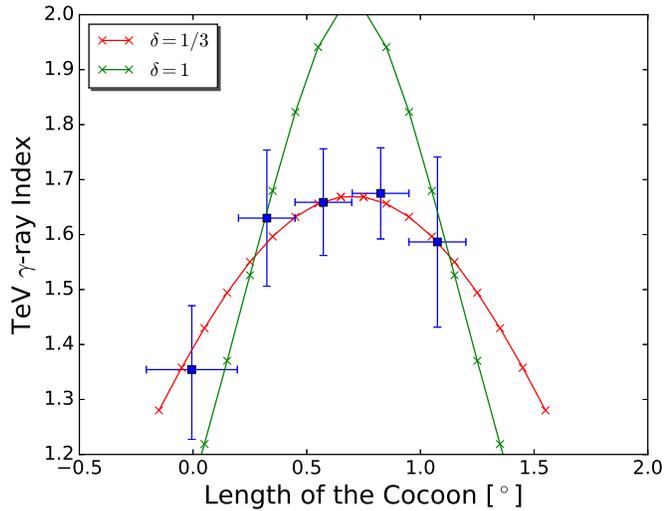}
		\caption{\textbf{Spatial distribution of $\gamma$-ray indices. The data are taken from \citet{2012A&A...548A..38A}.}}
		\label{fig:indices}
	\end{figure}
\end{center}

\begin{center}
\begin{figure}[H]
\includegraphics[scale=0.48]{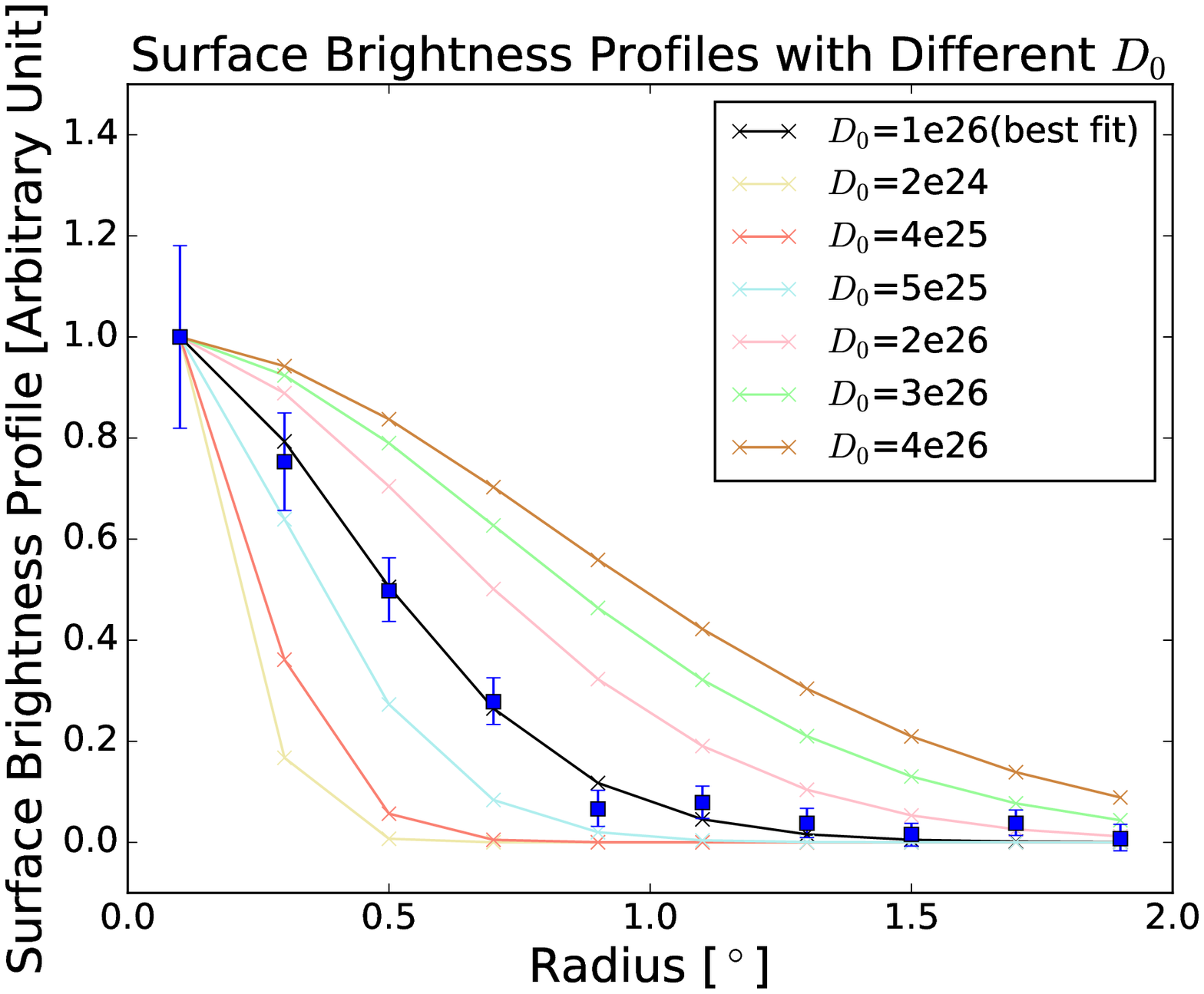}
\includegraphics[scale=0.48]{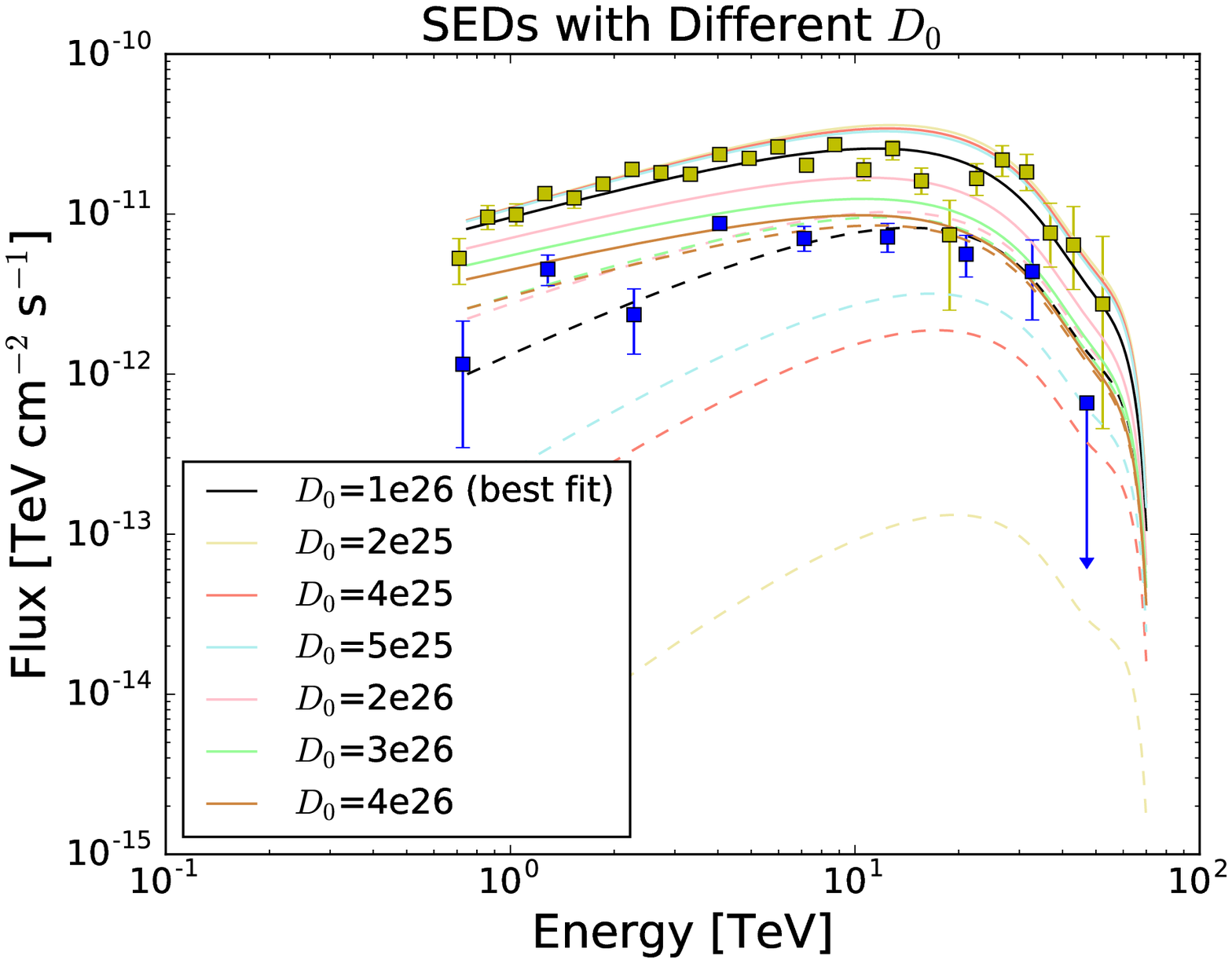}
\caption{Dependence of the model fit on the diffusion coefficient for $\alpha=1.7$. {The solid lines in the right panel represent the spectra of the inner region, and the dashed lines represent the spectra of the outer region.}}
\label{fig:general}
\end{figure}
\end{center}

 {\autoref{fig:elec} shows the spatially integrated electron distribution of the TeV {nebula} for parameters given in \autoref{tab:par}. The spectrum is cut off at $\sim 70$ TeV due to the radiative cooling.} \autoref{fig:Rings} shows the spectral fit to the inner and outer rings (left panel) and fit to the normalized brightness profile (right panel). {The SED can be fit well with $\alpha=1.7$, which is in agreement with the radio spectrum of the ERN (Paper II). {Since the magnetic field is weak in the X-ray cocoon, the synchrotron lifetime of 10\,TeV electrons is $ \approx 3\times 10^4(B/6\mu\textup{G})^{-2}$\,yr \citep{2011ApJ...743L...7H}, larger than the age of the Vela SNR, therefore the corresponding \textit{Fermi} $\gamma$-ray spectrum \citep{2018A&A...617A..78T} is a single powerlaw, consistent with theoretical expectations for $\gamma$-ray-emitting electrons with a power-law distribution. The spectrum of the outer ring is slightly harder than the inner ring due to the increase of the diffusion coefficient with energy. The spectrum of the outer ring is slightly harder than the inner ring due to the increase of the diffusion coefficient with energy.} In \autoref{fig:indices}, we show the spatial distribution of the $\gamma$-ray indices for different $\delta$. The $\gamma$-ray indices are calculated for nine annuli centered in the central position of the cocoon, with the same width of $0.1^\circ$. It can be seen that, for $\delta=1$, the $\gamma$-ray indices have apparent spatial variation, whereas for $\delta=1/3$, the spatial distributions of $\gamma$-ray indices are consistent with the indices extracted from sectors along the cocoon.} {The dependence of the TeV $\gamma$-ray brightness profile and spectra on the diffusion coefficient is plotted in \autoref{fig:general}.} It shows that the diffusion coefficient is well constrained ($D_0 \approx 1 \times10^{26}$\,cm$^{2}$\,s$^{-1}$) by the observed spectra and brightness profile. 

\begin{center}
\begin{deluxetable}{p{2.5cm}cc}
\tabletypesize{\footnotesize}
\tablecaption{Fitting Parameters\label{tab:par}}
\tablewidth{0pt}
\tablehead{
\\
Parameter     & Quantity \\
}
\startdata
$T_\textup{age}$ (yr)                        & 12000 \\
$Q_\textup{0}$ (cm$^{-3}$ s$^{-1}$)          & $6.7 \times 10^{36}$\\
$d_\textup{Vela}$ (pc)                       & 287\\
$\tau_\textup{s}$ (yr)					 & 7500 \\
$B_\textup{eff}$    ($\mu$G)	            & 6\\
$D_0$	  (cm$^2$ s$^{-1}$)			 & $10^{26}$\\
\hline
$\alpha$                                   & 1.7 \\
\enddata
\end{deluxetable}
\end{center}

\section{discussion and conclusions}
The discovery of a high-energy spectral cutoff in combination with a recent detection of a very hard GeV spectrum indicate that the $\gamma$-ray emission of the TeV nebula of the Vela X PWN is dominated by {electrons with energies of tens of TeV} via the IC process. The weak magnetic field strength derived from X-ray observations {and the high-energy cutoff are consistent with the radiative cooling of TeV electrons injected upon the formation of the cocoon, and the small total energy in the TeV nebula is consistent with the impulsive injection.} In this paper, {with} a simple diffusion and radiative loss model, {we demonstrate} that in the scenario {in which} the TeV cocoon formed via interaction of the PWN with the reverse shock several thousand years ago, the $\gamma$-ray spectra and radial brightness profile of the TeV nebula can be {reproduced.} Since the magnetic field strength is constrained by the injection time and {the electron index $\alpha$ can be constrained by the radio SED of the ERN}, the diffusion coefficient $D_0$ is the only key parameter that can be well constrained by the brightness profile and radial spectral change of the TeV nebula. {The diffusion coefficient of TeV electrons and positrons is thus determined to be $1 \times 10^{26}$\,cm$^{2}$\,s$^{-1}$ for 10\,TeV electrons, which is more than three orders of magnitude lower than the typical value in the ISM.} Our results, in combination with an early result from HAWC observations of the Geminga PWN suggest that slow diffusion {might be} common in TeV PWNe. 

{At 10\,TeV, the diffusion coefficient is almost the same as the Bohm diffusion ($8.3 \times 10^{25}$ cm$^2$ s$^{-1}$) in a magnetic field of $4\,\mu$G, which implies sub-Bohm diffusion at even higher energies. Since $r_\textup{dif} \sim 2\sqrt{DT_\textup{age}}\sim6$\,pc for 100\,TeV electrons, such slow diffusion could hardly transport the positrons below 100\,TeV out of the Vela SNR.} {One possible explanation for such a slow diffusion in the cocoon is efficient trapping of electrons in a magnetic mirror. Such trapping needs to be extended to the whole TeV nebula. Multiwavelength observations can be used to probe the magnetic field structure \citep{2012A&A...548A..38A}}. More detailed exploration of such a scenario will be presented in a future paper.

The energy of high-energy electrons inside the cocoon is estimated to be $E_\textup{cocoon}\sim 1.5 \times 10^{46}$\,erg \citep{2010ApJ...713..146A}, which is extremely low, {but much higher than the energy of $\sim 10^{44}$\,erg carried by the magnetic field}. The cocoon is expected to form due to the reverse shock-PWN interaction \citep{2018ApJ...865...86S}, hence the bulk of the pulsar spin-down energy (and bulk of TeV electrons) is deposited into the ERN. The SED of the ERN can be explained by intense radiation loss in the reverberation phase {for the interaction of the nebula with reverse shock}; the interpretation of the broadband spectrum of the ERN is delegated to an ensuing paper (Paper II). Moreover, if the magnetic field strength in the ERN is high, the soft GeV spectrum may be attributed to hadronic processes.


Slow diffusion around SNRs has been suggested for years \citep[see e.g.,][]{2009ApJ...707L.179F,2010sf2a.conf..313G,2010MNRAS.409L..35L,2012ApJ...745..140Y}. Theoretical works also show that cosmic rays may diffuse slowly due to self-generation waves and wave-wave turbulent cascading from a scale, which is in concordance with the size of the SNR \citep[see e.g.,][]{2012PhRvL.109f1101B}, and may also work inside PWNe {with weak magnetic fields}.

\acknowledgments
{We thank the anonymous referee for the helpful comments.} This work is supported by the National Key R\&D Program of China under grants 2018YFA0404203, 2015CB857100, and 2017YFA0402600, NSFC under grants 11773014, 11633007, 11851305, U1738122, and 11761131007, and the International Partnership Program of Chinese Academy of Sciences under grant 114332KYSB20170008.


\begin{thebibliography}{} 

\bibitem[Abdo et al.(2010)]{2010ApJ...713..146A} Abdo, A.~A., Ackermann, M., Ajello, M., et al.\ 2010, \apj, 713, 146 


\bibitem[Abdollahi et al.(2017)]{2017PhRvD..95h2007A} Abdollahi, S., Ackermann, M., Ajello, M., et al.\ 2017, \prd, 95, 082007 


\bibitem[Abeysekara et al.(2017)]{2017Sci...358..911A} Abeysekara, A.~U., Albert, A., Alfaro, R., et al.\ 2017, Science, 358, 911 


\bibitem[Abramowski et al.(2012)]{2012A&A...548A..38A} Abramowski, A., Acero, F., Aharonian, F., et al.\ 2012, \aap, 548, A38 


\bibitem[Accardo et al.(2014)]{2014PhRvL.113l1101A} Accardo, L., Aguilar, M., Aisa, D., et al.\ 2014, Physical Review Letters, 113, 121101 


\bibitem[Ackermann et al.(2012)]{2012PhRvL.108a1103A} Ackermann, M., Ajello, M., Allafort, A., et al.\ 2012, Physical Review Letters, 108, 011103 


\bibitem[Adriani et al.(2013)]{2013PhRvL.111h1102A} Adriani, O., Barbarino, G.~C., Bazilevskaya, G.~A., et al.\ 2013, Physical Review Letters, 111, 081102 

\bibitem[Aguilar et al.(2019)]{2019PhRvL.122d1102A} Aguilar, M., Ali Cavasonza, L., Ambrosi, G., et al.\ 2019, Physical Review Letters, 122, 041102 

\bibitem[Aharonian et al.(2006)]{2006A&A...448L..43A} Aharonian, F., Akhperjanian, A.~G., Bazer-Bachi, A.~R., et al.\ 2006, \aap, 448, L43 


\bibitem[Aschenbach et al.(1995)]{1995Natur.373..587A} Aschenbach, B., Egger, R., \& Tr{\"u}mper, J.\ 1995, \nat, 373, 587 


\bibitem[Atoyan et al.(1995)]{1995PhRvD..52.3265A} Atoyan, A.~M., Aharonian, F.~A., \& V{\"o}lk, H.~J.\ 1995, \prd, 52, 3265 

\bibitem[Bao \& Chen(2019)]{paperII} Bao, Y., \& Chen, Y.\ 2019, ApJ, accepted (Paper II)


\bibitem[Blasi et al.(2012)]{2012PhRvL.109f1101B} Blasi, P., Amato, E., \& Serpico, P.~D.\ 2012, Physical Review Letters, 109, 061101 




\bibitem[DAMPE Collaboration et al.(2017)]{2017Natur.552...63D} DAMPE Collaboration, Ambrosi, G., An, Q., et al.\ 2017, \nat, 552, 63 

\bibitem[de Jager et al.(2008)]{2008ApJ...689L.125D} de Jager, O.~C., Slane, P.~O., \& LaMassa, S.\ 2008, \apjl, 689, L125 

\bibitem[Dodson et al.(2003a)]{2003ApJ...596.1137D} Dodson, R., Legge, D., Reynolds, J.~E., \& McCulloch, P.~M.\ 2003a, \apj, 596, 1137 


\bibitem[Dodson et al.(2003b)]{2003MNRAS.343..116D} Dodson, R., Lewis, D., McConnell, D., \& Deshpande, A.~A.\ 2003b, \mnras, 343, 116 





\bibitem[Fang et al.(2018)]{2018ApJ...863...30F} Fang, K., Bi, X.-J., Yin, P.-F., \& Yuan, Q.\ 2018, \apj, 863, 30 

\bibitem[Frail et al.(1997)]{1997ApJ...475..224F} Frail, D.~A., Bietenholz, M.~F., \& Markwardt, C.~B.\ 1997, \apj, 475, 224 

\bibitem[Fujita et al.(2009)]{2009ApJ...707L.179F} Fujita, Y., Ohira, Y., Tanaka, S.~J., \& Takahara, F.\ 2009, \apjl, 707, L179 


\bibitem[Gabici et al.(2010)]{2010sf2a.conf..313G} Gabici, S., Casanova, S., Aharonian, F.~A., \& Rowell, G.\ 2010, SF2A-2010: Proceedings of the Annual meeting of the French Society of Astronomy and Astrophysics, 313 


\bibitem[Gaensler \& Slane(2006)]{2006ARA&A..44...17G} Gaensler, B.~M., \& Slane, P.~O.\ 2006, \araa, 44, 17 


\bibitem[Hinton et al.(2011)]{2011ApJ...743L...7H} Hinton, J.~A., Funk, S., Parsons, R.~D., \& Ohm, S.\ 2011, \apjl, 743, L7 


\bibitem[Hooper et al.(2009)]{2009JCAP...01..025H} Hooper, D., Blasi, P., \& Serpico, P.~D.\ 2009, \jcap, 1, 025 

\bibitem[Hooper et al.(2017)]{2017PhRvD..96j3013H} Hooper, D., Cholis, I., Linden, T., \& Fang, K.\ 2017, \prd, 96, 103013 


\bibitem[Huang et al.(2018)]{2018ApJ...866..143H} Huang, Z.-Q., Fang, K., Liu, R.-Y., \& Wang, X.-Y.\ 2018, \apj, 866, 143 


\bibitem[Li \& Chen(2010)]{2010MNRAS.409L..35L} Li, H., \& Chen, Y.\ 2010, \mnras, 409, L35 




\bibitem[Malkov et al.(2013)]{2013ApJ...768...73M} Malkov, M.~A., Diamond, P.~H., Sagdeev, R.~Z., Aharonian, F.~A., \& Moskalenko, I.~V.\ 2013, \apj, 768, 73 

\bibitem[Markwardt \& {\"O}gelman(1995)]{1995Natur.375...40M} Markwardt, C.~B., \& {\"O}gelman, H.\ 1995, \nat, 375, 40 

\bibitem[Porter et al.(2008)]{2008ApJ...682..400P} Porter, T.~A., Moskalenko, I.~V., Strong, A.~W., Orlando, E., \& Bouchet, L.\ 2008, \apj, 682, 400 


\bibitem[Recchia et al.(2018)]{2018arXiv181107551R} Recchia, S., Gabici, S., Aharonian, F.~A., \& Vink, J.\ 2018, arXiv:1811.07551 

\bibitem[Reynolds et al.(2017)]{2017SSRv..207..175R} Reynolds, S.~P., Pavlov, G.~G., Kargaltsev, O., et al.\ 2017, \ssr, 207, 175 

\bibitem[Shi \& Liu(2019)]{2019MNRAS.tmp..670S} Shi, Z.-D., \& Liu, S.\ 2019, \mnras,  

\bibitem[Slane et al.(2018)]{2018ApJ...865...86S} Slane, P., Lovchinsky, I., Kolb, C., et al.\ 2018, \apj, 865, 86 


\bibitem[Staszak \& VERITAS Collaboration(2015)]{2015ICRC...34..411S} Staszak, D., \& VERITAS Collaboration 2015, 34th International Cosmic Ray Conference (ICRC2015), 34, 411 


\bibitem[Tibaldo et al.(2018)]{2018A&A...617A..78T} Tibaldo, L., Zanin, R., Faggioli, G., et al.\ 2018, \aap, 617, A78 


\bibitem[Xi et al.(2018)]{2018arXiv181010928X} Xi, S.-Q., Liu, R.-Y., Huang, Z.-Q., et al.\ 2018, arXiv:1810.10928 


\bibitem[Yan et al.(2012)]{2012ApJ...745..140Y} Yan, H., Lazarian, A., \& Schlickeiser, R.\ 2012, \apj, 745, 140 


\end{thebibliography}
\end{document}